**Spontaneous Cyclogenesis without Radiative and Surface-Flux Feedbacks**

Argel Ramírez Reyes (1) and Da Yang* (1, 2)

Affiliations:

1.  University of California, Davis

2.  Lawrence Berkeley National Laboratory

*Contact information: dayang@ucdavis.edu

**Abstract**:

Tropical cyclones (TCs) are among the most intense and feared storms in the world. What physical processes lead to cyclogenesis remains the most mysterious aspect of TC physics. Here, we study spontaneous TC genesis in rotating radiative-convective equilibrium using cloud-resolving simulations over an $f$-plane with constant sea-surface temperature. Previous studies proposed that spontaneous TC genesis requires either radiative or surface-flux feedbacks. To test this hypothesis, we perform mechanism-denial experiments, in which we switch off both feedback processes in numerical simulations. We find that TCs can self-emerge even without radiative and surface-flux feedbacks. Although these feedbacks accelerate the genesis and impact the size of the TCs, TCs in the experiments without them can reach similar intensities as those in the control experiment. We show that TC genesis is associated with an increase in the Available Potential Energy (APE); and that convective heating dominates the APE production. Our result suggests that spontaneous TC genesis may result from a cooperative interaction between convection and circulation, and that radiative and surface-flux feedbacks accelerate the process. Furthermore, we find that increasing the planetary rotation favors spontaneous TC genesis.

## 1. Introduction

A tropical cyclone (TC) is a rapidly cyclonically rotating storm system that typically forms over a tropical ocean. It is often characterized by a center of anomalously low surface pressure, a closed low-level atmospheric circulation, a warm core, and a spiral arrangement of thunderstorms. TCs are among the most intense and feared storms of the world, with torrential rains that can last for $O(1 \text{ week})$ and destructive winds that span over $O(500 \text{ km})$. In spite of extensive theoretical developments and the ever-advancing observing capabilities, TC genesis remains as the most mysterious aspect of TC physics [see Emanuel (2018) for a comprehensive review].

Observational studies suggested that TCs often form from a pre-existing mid-level cyclonic vortex (McBride and Zehr 1981; Davidson et al. 1990; Bartels and Maddox 1991; Laing and Fritsch 1993; Velasco and Fritsch 2012). Recent studies observed that a closed "pouch" associated with a tropical wave protects a region of deep convection from the intrusion of dry air, favoring the formation of a surface-concentrated vortex that then intensifies to a TC (Dunkerton et al. 2009;

Wang et al. 2010; Raymond and López Carrillo 2011; Wang 2012; Smith et al. 2015). Although these studies have provided many insights, the high degree of complexity in the real atmosphere makes it difficult to distinguish what physical processes are essential to TC genesis. Here, idealized modeling studies are ideal complements to the observational studies.

Recent work has shown that TCs can spontaneously develop in rotating radiative-convective equilibrium (RRCE) simulations using both cloud-resolving models (CRMs) and general circulation models (GCMs) (Bretherton et al. 2005; Nolan et al. 2007b; Held and Zhao 2008; Khairoutdinov and Emanuel 2013; Zhou et al. 2013; Shi and Bretherton 2014; Chavas and Emanuel 2014; Reed and Chavas 2015; Davis 2015; Merlis et al. 2016; Wing et al. 2016; Muller and Romps 2018; Chavas and Reed 2019; Merlis and Held 2019; Carstens and Wing 2020). These studies suggest that initial disturbances may help TC genesis in the real atmosphere, but they are *not* essential.

Spontaneous TC genesis is considered as $f$-plane convective self-aggregation. Convective self-aggregation is a phenomenon in which large-scale circulations and convective organization can self-emerge over uniform sea-surface temperatures (SSTs) and boundary conditions (Bretherton et al. 2005; Muller and Held 2012; Arnold and Randall 2015; Wing et al. 2018; Yang 2018a,b, 2019). This process is associated with increasing variance of moist static energy (MSE, an approximation for the total energy contained in a moist air parcel), and increasing eddy available potential energy (APE, defined in section 2). Wing et al. (2016) analyzed the MSE variance budget in spontaneous TC genesis and showed that radiative and surface flux feedbacks help increase MSE variance, consistent with non-rotating self-aggregation processes. In addition, the authors removed the radiative feedback by horizontally homogenizing radiative cooling rates and confirmed the results of Frisius (2006), who showed that spontaneous TC genesis was delayed in these conditions. Muller and Romps (2018) further showed that removing surface-flux feedbacks significantly delays TC genesis and reduces the TC strength at the mature stage. These mechanism-denial experiments are consistent with the MSE analysis, showing that radiative and surface-flux feedbacks contribute to increase MSE variance, favoring TC genesis. As the MSE variance increase is associated with radiative and surface-flux feedbacks, the MSE analyses seem to suggest that spontaneous TC genesis would not occur in the absence of radiative and surface-flux feedbacks. However, as MSE is approximately conserved for an undiluted air parcel undergoing moist adiabatic processes (Romps 2015), the MSE analysis does not explicitly show the role of convective heating.

Evaporation of water from the ocean surface constitutes the most important source of energy for TCs (Emanuel 2003, 2018). This energy, later released by condensation of water vapor, is key to TC development and maintenance: Convective heating can generate APE, which can subsequently convert to kinetic energy, providing energy for TC genesis and maintenance (Nolan et al. 2007a). Generation of APE is also associated with the development of a warm core, which is necessary to sustain the vortex in hydrostatic and gradient wind balances. Although an APE-centric framework has been widely used to understand convectively coupled tropical circulations (including non-rotating convective self-aggregation and TC genesis in axisymmetric models) (Anthes and Johnson 1968; Emanuel et al. 1994; Nolan et al. 2007a; Veiga et al. 2008; Kuang 2008; Wong et al. 2016; Yang 2018a), it has not been applied to understand spontaneous TC genesis. In contrast to MSE, increases in APE are led by APE production due to convection, radiation, and surface fluxes (*e.g.*, Yang 2018a). Therefore, an analysis of the APE highlights the role of convection, complementing the MSE analysis.

In this paper, we combine the analysis of APE with mechanism-denial CRM simulations to address the question of what processes contribute to the spontaneous TC genesis, and what is the minimum recipe. This study aims to extend our understanding of spontaneous TC genesis, moist convection, and convective organization by expanding the parameter space used in previous studies to attack the question of whether convection can or cannot aggregate in the form of a TC without the action of radiative and surface-flux feedbacks. This study is also relevant to the study of planetary atmospheres in which a larger parameter space of rotation rates can be expected. We describe our research methods in Section 2, present simulation results in Section 3, and expose the APE analysis in Section 4. In section 5 we show the sensitivity of our results to changes of the Coriolis parameter, resolution, initial conditions, temperature and parametrizations of radiation and microphysics. We conclude and discuss the implications of our findings in Section 6.

## 2. Methods

### 2.1 Available potential energy

APE is the amount of gravitational potential energy that can be transformed into kinetic energy by lowering the center of mass of the atmosphere (Vallis 2017). APE is, therefore, an energy reservoir for atmospheric circulations (Lorenz 1955). APE can be computed as the difference between the potential energy in the fluid and the minimum potential energy possible for the same system after an adiabatic rearrangement of mass. In a *dry* APE framework (one that externalizes convective heating), the APE of an anelastic atmosphere is given by (e.g., Yang 2018a)

$$APE = \frac{1}{2} \int_0^z \rho_0(z') \frac{\overline{b^2}(z')}{N^2(z')} dz', \tag{1}$$

where $b = b(x, y, z, t)$ is buoyancy, $\rho_0(z)$ represents the reference density, $N^2(z) = g\frac{1}{\theta_v}\frac{\partial \overline{\theta_v(z)}}{\partial z}$ is the Brunt-Väisälä frequency squared (a measure of stratification), $z$ denotes the vertical coordinate, $\theta_v = T(1 + \epsilon q)(\frac{p_0}{p})^{\left(\frac{R_d}{c_{pd}}\right)}$ is the virtual potential temperature where T is the temperature, q the specific humidity, $p_0$ and $p$ are pressure and reference pressures, $R_d$ and $c_{pd}$ are the gas constant and specific heat capacity at constant pressure for dry air, respectively, and $\epsilon = \frac{M_{air}}{M_{water}} - 1 = 0.61$. In here and in the following, the horizontal bar represents horizontal averaging over the region considered and a prime denotes departures from said average.

To obtain an evolution equation for the APE, we consider the buoyancy equation

$$b = g\frac{\theta_v'}{\overline{\theta_v}}, \tag{2.a}$$

$$\partial_t b + u\partial_x b + v\partial_y b + wN^2 = S_b, \tag{2.b}$$

where $t$ is time, $u, v$ and $w$ are the components of the velocity vector, b is the buoyancy computed in a moist atmosphere (equation 6.1.5 in Emanuel 1994), and $S_b$ represents buoyancy sources, including convective heating, radiative cooling, and surface fluxes. Equation (2.a) differs from SAM's formulation of buoyancy in that it does not consider the effect of condensate loading. However, this widely used approximation should not impact the computation of the slow-varying component of APE. Using (1) and (2), Yang (2018a) obtained an evolution equation for the slowly varying component of the APE:

$$\overbrace{\frac{1}{2}\int_0^z \frac{\rho_0}{N^2}\ \partial_t \widetilde{b}^2\ dz'}^{\partial_t \overline{APE}} = \overbrace{\int_0^z \frac{\rho_0}{N^2}\widetilde{\overline{bS_b}}\ dz'}^{Production} - \overbrace{\int_0^z \frac{\rho_0}{N^2}(\widetilde{\overline{bu\partial_x b}}\ +\ \widetilde{\overline{bv\partial_y b}}\ )\ dz'}^{Advection} - \overbrace{\int_0^z \rho_0\widetilde{\overline{b\widetilde{w}}}\ dz'}^{Conversion}, (3)$$

where the first term on the right-hand side is the APE production term, the second and third are the advection terms, and the last term is the conversion to kinetic energy. The tilde represents the slow-varying component. We focus our study on the slow-varying component of APE which we expect to be associated with TCs. According to Yang (2018a), buoyancy sources are computed by

$$S_b = g\frac{S_\theta}{\overline{\theta}} + g\frac{\epsilon S_q}{1 + \epsilon \overline{q}}\ , \tag{4}$$

where $\theta$ is potential temperature, $q$ is specific humidity, $g$ is the acceleration of gravity, $S_q$ and $S_\theta$ are sources of humidity and heating, respectively. It is important to note from (3) that the production of APE and growth of its associated circulations occurs when the product $bS_b$ is positive (when buoyancy and buoyancy sources are positively correlated); when the product is negative, APE is reduced and its associated circulations decay (Emanuel et al. 1994; Yang 2018a). Because anomalies in convective heating, surface heat fluxes and radiative heating are buoyancy sources, they contribute to the APE production.

2.2 Model and experiment setup

We perform rotating RCE simulations over an *f*-plane using the System for Atmospheric Modeling, SAM, version 6.10.10 (Khairoutdinov and Randall 2003). SAM solves the anelastic system of equations and prognoses liquid water and ice moist static energy, total nonprecipitating water (vapor + cloud water + cloud ice), and total precipitating water (rain + snow + graupel). SAM has been widely used to study tropical convection (e.g Bretherton et al., 2005; Khairoutdinov & Emanuel, 2013; Muller & Held, 2012; Wing et al., 2016; Yang, 2018a). The radiation scheme is that of the Community Atmosphere Model 3 (Collins et al. 2004), which computes long-wave and short-wave radiative heating rate at each level of each column every 15 time steps; the sub grid-scale scheme is the SAM Smagorinsky scheme, which is the stationary version of a prognostic scheme based on turbulent kinetic energy, the implementation of which is described in (Deardorff 1980); and the microphysics scheme is the SAM single-moment microphysics (Khairoutdinov and Randall 2003), which computes the sum of nonprecipitating liquid and ice water and water vapor mixing ratio and the total precipitation mixing ratio. The surface fluxes are computed with bulk formulae with exchange coefficients computed using the Monin-Obukhov theory, with code adapted from NCAR's Community Climate Model 3 (Kiehl et al. 1996). SAM does not have a parameterization scheme for the boundary layer.

The main four experiments are summarized in Table 1, and they have a shared configuration consisting of a doubly periodic horizontal domain of 1024 km × 1024 km with a horizontal grid spacing of 2 km. The vertical domain is of 34.8 km, and the vertical grid-spacing is 50 m from $z = 0$ m to $z = 1050$ m and then it increases gradually until it reaches 600 m at $z = 3000$ m. The integration timestep is 10 s, but it decreases when needed to prevent numerical instabilities. The simulations are run for 100 days starting with an RCE profile produced as the mean potential temperature and humidity sounding from the last 20 days of a smaller 2D simulation. The sounding profiles are the same as those used in (Yang 2018a). We save 3D variables every two hours and 2D variables every hour. Following Khairoutdinov and Emanuel (2013), we use a constant Coriolis parameter $f = 4.97 \times 10^{-4}$ s$^{-1}$. This value corresponds to 10 times the Coriolis parameter at 20° latitude. The large $f$ helps shrink the horizontal scale of TCs and allows us to simulate TCs in a small domain. Although intuitively a higher Coriolis parameter would also accelerate the TC genesis process (e.g., by increasing vorticity due to stretching), previous studies have not found a robust relation between time to genesis and rotation rate in $f$-plane simulations (Nolan et al. 2007b; Carstens and Wing 2020), thus we don't expect this choice to have a leading order impact in the time to genesis in our simulations. Similar values have been used in other studies of spontaneous TC genesis (Chavas and Emanuel 2014; Cronin and Chavas 2019). The choices of $f$, domain size and resolution stem from a necessary compromise: using a realistic value of $f$ with 2-km grid-spacing would require a much larger domain that becomes computationally prohibitive. On the other hand, a grid spacing of 2-km is desirable to have a good representation of some aspects of convection, for example in-cloud downdrafts and updrafts and mid-troposphere entrainment (Nolan et al. 2007b). For consistency with previous studies of non-rotating convective aggregation in RCE using doubly periodic domains (Muller and Held 2012; Jeevanjee and Romps 2013; Yang 2018a), the horizontally averaged wind speed is nudged to zero at all levels with a timescale of two hours. This prevents the emergence of wind shear, which in a rotating setup adds complexity to study TC formation (Nolan et al. 2007b). However, nudging the wind speed does not strongly affect our results (see Appendix 1). The sea surface temperature is held at 300 K. The diurnal cycle is turned off and the solar insolation is constant set to $685 \, \mathrm{W \, m^{-2}}$, with a solar zenith angle of 51.7 degrees, similar to the value used in (Tompkins and Craig 1998). The model has a rigid lid, but SAM applies Newtonian damping to all the prognostic variables in the upper third of the domain to prevent gravity wave reflection.

## 2.3 Mechanism-denial experiments: turning off feedbacks

We perform mechanism-denial experiments to investigate the sensitivity of TC genesis to the radiative and surface-flux feedbacks. The radiation-moisture feedback appears when the atmosphere is organized in moist, convecting regions and dry, subsiding regions. In this case, radiative cooling leads to subsidence in the dry area, which promotes more radiative cooling and further dries the atmosphere. To disable this feedback, we follow Muller and Romps (2018) in substituting the radiative cooling rate at each grid point with its horizontal average, as commonly done in similar studies of spontaneous TC genesis (Wing et al. 2016) and convective self-aggregation (Bretherton et al. 2005; Yang 2018a). Substituting radiative fluxes by their horizontal mean removes horizontal radiative heating anomalies, so radiation cannot contribute to the generation of APE, effectively removing the feedback.

The surface-flux feedbacks comprise two competing effects. The buoyancy fluxes from the sea-surface to the atmosphere depend on wind speed and sea-air differences in moisture and temperature through the latent and sensible heat fluxes. For example, the latent heat fluxes are computed as $LHF = \rho C_E L_v U (q_{T_s}^* - q_v)$, where $\rho$ is the density of the air, $C_E$, an exchange coefficient, $L_v$, the latent heat of vaporization of water, U the air wind speed near the surface and $q_{T_s}^*, q_v$, the saturation specific humidity and specific humidity, respectively. An enhanced wind speed near the surface promotes moisture fluxes to the atmosphere, which in turn increases the overturning circulation and enhances surface wind speeds, closing the feedback loop. On the other hand, regions of enhanced air humidity have a decreased humidity difference. This acts to decrease the surface fluxes, weakening the convection and resulting in a negative feedback. A similar argument holds for the sensible heat flux, $SHF = \rho C_H C_p U(T_s - T_a)$ with $C_p, C_H, T_s$ and $T_a$ are the dry air specific heat capacity at constant pressure, a transfer coefficient and the temperatures of the ocean surface and atmosphere, respectively. Following Muller and Romps (2018), we remove the surface-flux feedbacks by replacing the surface heat fluxes at each grid point with their horizontally averaged value, as in previous studies (Bretherton et al. 2005; Yang 2018a). In this case, surface fluxes do not contribute to APE production.

We will present four sets of simulations: Control, where the interactive radiation and surface fluxes are not altered; HomoRad, where the radiative feedback is turned off; HomoSfc, where the surface-flux feedbacks are removed, and HomoAll where both radiative and surface-flux feedbacks are turned off (see Table 1).

### 2.4 Sensitivity experiments

We perform a suite of experiments to test the sensitivity of our results to changes in resolution, Coriolis parameter, initial conditions, sea-surface temperature and choices of radiation and microphysics schemes (Table 2). In the following we summarize the design of these sensitivity simulations, where we keep the domain and the rest of configurations unaltered with respect to the experiments in Table 1.

   a) Initial condition: We create two additional initial profiles that differ from each other and from the original sounding used to initialize the simulations of Table 1. Using these initial profiles, we run two 50-day simulations with each of the Control, HomoRad, HomoSfc and HomoAll configurations. To construct the two different initial conditions, we add random noise in the first five levels of the original potential temperature sounding, as done by Wing et al. (2016). The temperature perturbation has an amplitude of 0.1 K in the first level and decreases linearly to 0.02K in the fifth level.

   b) Horizontal resolution: We perform a 30-day simulation with horizontal grid-spacing of 1 km in the HomoAll configuration to explore the sensitivity of the results to changes in resolution.

c) Coriolis parameter: We run 50-day simulations with $f = 3 \times 10^{-4}\,\mathrm{s}^{-1}$ and $f = 1 \times 10^{-4}\,\mathrm{s}^{-1}$ in the HomoAll configuration, respectively.

d) Microphysics: We conduct a 50-day HomoAll simulation but with the Thomson microphysics scheme (Thompson et al. 2008) which is a one moment bulk parameterization that predicts the mixing ratios of cloud water, rain, cloud ice, snow and graupel, and also predicts the number concentration of cloud ice

e) Radiation: We run a 50-day HomoAll simulation but with the RRTM radiation scheme (Mlawer et al. 1997).

2.4 APE computation and TC composite.

We compute APE and APE budgets from Equations 1, 2 and 3. Convective heating is not a standard output of SAM, which solves the conservation law of MSE. Therefore, to compute the contribution of convective heating to APE production, we diagnose convective heating from the buoyancy equation: we first calculate left-hand-side terms in (Equation 2.b); we then calculate convective heating as the residual of (Equation 2.b)—the left-hand-side terms minus radiative cooling and surface buoyancy fluxes. We compute the convective heating from the 3D output (every two hours) for the whole simulation. After computing convective heating, we have all the variables needed to compute APE and the APE budget using equation 2. Following Yang (2018a), to focus on the slow-varying component of APE evolution, we use a moving mean filter in the variables that enter the APE and APE budget computation. The moving mean filter has a window size of 30km in the horizontal directions and 5 days in the time dimension, but other averaging windows show similar results.

To characterize basic TC features, we make an "average TC" using the last 50 days of each of the simulations shown in Table 1 (or 600 output times to match the bi-hourly frequency of the rest of the analysis which uses 3D data). Similar to Zhou et al. (2017), we identify TCs as the point of minimum surface pressure withing regions of anomalously low surface pressure. We first find the departure of surface pressure from its horizontal mean ($P'_{sfc}(x,y,t) = P_{sfc}(x,y,t) - \overline{P_{sfc}}(t)$, where the overbar denotes the horizonal mean). Because the scale of a tropical cyclone is of $O(500$ km), we smooth this pressure anomaly with a moving median filter to reduce the smaller scale features. We use a window size of 20 km for smoothing and our results are robust over different choices of the window sizes. We then identify contiguous regions of pressure anomaly less than $-9$ hPa. Next, we find the point of minimum pressure within this region, which we record as a TC. Identifying TC centers as a point of minimum pressure has been used in other studies (e.g. Reed and Chavas 2015), so this method is adequate to build our composites. Having obtained the time and location of the TC centers (the minima of surface pressure perturbation), we then create the TC composite by aligning the centers of the identified TCs and taking the time average only over the time steps with identified TCs. This gives a composite of TC-associated variables, including surface pressure, surface winds, air temperature, and others. We then compute the azimuthal average of all the quantities in radial bins of 2km width starting at r = 1km as done by Cronin and Chavas (2019). This process yields radial profiles of the characteristic features of the TC composite.

## 3 Results

TCs self-emerge in all four simulations. Movie S1 shows the time evolution of surface pressure and wind speed from a homogeneous state. Figure 1 shows contours of surface winds and surface pressure over a snapshot at day 70 for the Control, HomoRad, HomoSfc and HomoAll experiments. All experiments simulate TC-like structures, featuring organized areas of low surface pressure (< 990 hPa to be noticeable in the Movie S1) and high wind speed (> $10 \text{ ms}^{-1}$ to be noticeable in the Movie S1), with a clearly defined eye region in the center. In the Control simulation, regions of enhanced wind speed and a center of low pressure occur after 7 days. By day 8, multiple centers of low pressure, and high wind speed have emerged. As time progresses, the low-pressure centers become centers of low wind speed surrounded by rapidly rotating wind, and a clear eye of low pressure surrounded by an eyewall of high wind speed is discernible by day 9. We consider at this point that TCs are developed. By day 15 we can observe up to 7 TCs in the domain with wind speeds as high as 50 m/s. The multiple TCs merge. After 33 days, we can only observe three TCs, and only two TCs of greater horizontal extent exist from day 68 till the end of the simulation.

In HomoRad, TC genesis is slower: the regions of enhanced wind speeds appear around day 10, and several clearly defined TCs are developed after 15 days. We observe up to 9 TCs coexisting in the same domain by day 16. Multiple TCs have merged by day 66, and only 3 TCs exist for the rest of the simulation. In HomoSfc, regions of low surface pressure appear by day 13, and one TC clearly formed by day 22. Notably, the closed region of maximum wind speed around the center is less well defined than in the previous to simulations, suggesting an important role of surface fluxes for maintaining the structure of the TCs. The genesis process is further delayed in HomoAll. The region of enhanced wind speeds can be first observed around day 22, and the first clearly defined TC is developed by day 38. During the last 20 days of the simulation, two TCs coexist. They both have similar intensities, but one is notably smaller than the other.

TCs without radiative and surface-flux feedbacks can reach the same intensity as those in the Control simulation. In Figure 1 and Movie S1, we observe TCs in all experiments, showing a *similar intensity*. Figure 2 shows the temporal evolution of maximum surface wind speed (Figure 2a) and minimum surface pressure (Figure 2b) in the domain of each experiment. The first 50 days of Figure 2 show a continuous line which is a mean of the 3-member ensemble (see Table 1 and Table 2) for each experiment, and the ribbon spans from the minimum to the maximum value of these 3 members. The next 50 days show only the values for the simulations described in Table 1. After an initial period of intensification, the maximum surface wind speed and minimum surface pressure in all the experiments oscillate around similar values. In the Control and HomoRad experiments, maximum surface wind speed and minimum surface pressure first reach a maximum and minimum value respectively, after which the maximum wind speed and minimum surface pressure oscillate around a slightly lower wind speed and higher surface pressure for the rest of the simulation. It is interesting to note that after the first 50 days, the HomoRad experiments achieve the strongest wind speed (above $50 \ \mathrm{ms^{-1}}$) and lowest surface pressure (around 920 hPa) among the 4 experiments of Table 1, suggesting TCs with the strongest intensity. However, during the first 50 days, the ensembles of Control and HomoRad overlap significantly. The fact that TCs in Control and HomoRad are of the same intensity to the leading order is consistent with previous studies (Muller and Romps 2018). In all four experiments, the maximum wind speed does not vary significantly after the first 50 days. The standard deviation of the maximum wind speed in the last 50 days is about 12% in control, 7% in HomoRad, 15% in HomoSfc, and 15% in HomoAll, suggesting that the system has achieved a statistical steady state. When the grid-spacing is reduced to 1-km in HomoAll, the intensity remains similar to that of the 2-km simulation (see Movie 2), suggesting that the intensity of the TCs converge with increased resolution.

Radiative and surface-flux feedbacks accelerate spontaneous TC genesis (Fig. 2). We identify TC genesis as the period when the maximum surface wind speed first reaches 33 m/s, similar to (Wing et al. 2016), and we mark this moment with a star on top of the line. This moment occurs around day 7.6 in Control, day 11.4 in HomoRad, day 15 in HomoSfc and day 35.5 in HomoAll. This occurs around the time the rotation is noticeable in the Movie S1: around day 8 in the Control simulation, day 10 in HomoRad, day 13 in HomoSfc, and day 33 HomoAll, showing that genesis is also slower when removing both feedbacks. The order of emergence of TCs is consistent across ensemble members (first in Control, second in HomoRad, third in HomoSfc and fourth in HomoAll). The acceleration of TC genesis by radiative and surface-flux feedbacks is consistent with the results of Muller and Romps (2018), Wing et al. (2016) and Zhang and Emanuel (2016).

TC composites show that the simulated TCs have horizontal structures that resemble observations of real TCs. Figure 3 shows the azimuthal average of surface wind speed (3a) and surface pressure (3b) of the composites. TCs in the four simulations are all characterized by a center of quiescent winds collocated with a well-defined eye of minimum surface pressure. Both the azimuthal averages of tangential and radial surface wind speeds have local extrema (minimum for the tangential component and maximum for the radial component) at the center of minimum pressure in the three cases. The *mean* tangential wind increases rapidly outward while the radial component decreases (Figure 3a). The *mean* tangential wind speed is around 24 m s$^{-1}$ at 40 km from the center in Control, 32 m s$^{-1}$ at 22 km from the center in HomoRad, 21 m s$^{-1}$ at 50 km in homoSfc, and 17 m s$^{-1}$ at about 20 km from the center in HomoAll. After reaching the maximum, the tangential wind speed then decreases with distance from the center, reaching 12 m s$^{-1}$ at 200 km in Control, at around 165 km in HomoRad and HomoSfc and at 100 km in HomoAll. This shows that the TCs in HomoAll have a smaller horizontal extent than the other three experiments. Our results are robust over different definitions on the size of the average storm, e.g., the radius of maximum azimuthally averaged tangential wind, or the radius where tangential wind reaches 12 m s$^{-1}$ (Chavas et al. 2016). The radial wind speed has a similar structure, reaching a minimum and then increasing with radius. Similarly, the surface pressure has a local minimum in the center of the TC, and it increases with distance (Figure 5b). The *mean* surface pressure minimum is 958 hPa in Control, 947 hPa in HomoRad, 972 hPa in HomoSfc and 977 hPa in HomoAll, and the ambient surface pressure is around 1000 hPa in all the experiments.

The vertical structures of the simulated TCs also are consistent with realistic TCs. Figure 4 shows azimuthal averages of tangential wind speed in shading and black contours of virtual potential temperature anomaly (4a-4d), and convective heating anomaly (4e-4h) with a horizontal red line denoting the radiative tropopause, computed using the last 50 days of simulations, as the height at which the horizontally averaged time-mean radiative cooling rate vanishes (Pierrehumbert 2010; Cronin and Chavas 2018; Seidel and Yang 2020). The tangential wind speed is positive, indicating cyclonically rotating wind throughout much of the troposphere (Figure 4a – Figure 4d). Above this cyclonic wind, we observe tangential wind in the opposite direction in the four experiments. Additionally, Figures 4a-4d, show a buoyant (warm) core in the center of each TC, indicated by the contours of virtual potential temperature anomaly. The warm core extends vertically to the tropopause, which is at 15.7 km in Control ,16 km in HomoRad, 14.9 in HomoSfc and 14.4 km in HomoAll.

Positive convective heating anomalies extend from near the surface to the tropopause and are more intense in the eyewall, a narrow region relatively close to the center, (Fig. 4e-4h). This region is partially collocated with the warm core of the simulated TCs. The deep heating structure in the troposphere captures latent heat release in convective storms with a maximum intensity in the middle troposphere, around 7.5 km in Control, around 8 km in HomoRad, around 6.5 km in HomoSfc and around 6 km in HomoAll. It is worth noting that all the simulations present negative convective heating anomaly at the center . The region of strong convective heating is wider in Control and in HomoSfc than in HomoRad and HomoAll. An interesting feature is that the convective heating in the HomoRad experiment is twice as strong as that in the rest of the simulations, and in this case the heating also coincides with a higher temperature perturbation than in the rest of the experiments due to the heating being concentrated in a thinner region than in the rest of the simulations. The spatial coincidence of positive convective heating and buoyancy anomalies seen in Figure 4, suggests positive APE production due to convection (as observed in section 2.1). To examine further this hypothesis, we now examine the time evolution of APE and APE budgets in the simulations.

## 4 Evolution of APE in the simulated TCs

TC development is associated with APE evolution. Figure 5a shows the time evolution of the total APE in the domain for the four experiments shown in Table 1. In all simulations, the APE grows initially with time (the genesis period) and reaches the first local maximum around days $12, 24, 28$ and $20$ for Control, HomoRad, HomoSfc and HomoAll, respectively. For reference, the time to genesis is marked with a star at the same position as in Figure 2. We can see that genesis is led by APE growth, and as time advances and more TCs appear and intensify, APE further increases. This suggests that diagnosing APE evolution may help understand TC genesis and intensification.

Figures 5b-5e show the APE budgets. In all simulations, convective heating dominates APE production, and the APE production by convective heating is mainly balanced by its conversion to kinetic energy $(-wb)$, whereas the contribution of radiative and surface fluxes is modest. These results hold, in particular, for Control, HomoRad and HomoSfc, where radiative and/or surface-flux feedbacks are active. The magnitude of the APE budgets is similar in all the simulations. This observation may help explain the occurrence of TCs in HomoAll after removing radiative and surface-flux feedbacks.

Spontaneous TC genesis without radiative and surface-flux feedbacks challenge the prevailing theory of spontaneous TC genesis. Previous studies regarded surface-flux or radiative feedbacks as essential ingredients in the spontaneous TC genesis process (Wing et al. 2018; Muller and Romps 2018). However, we found that TCs can self-emerge without radiative and surface-flux feedbacks (Figures 1 and 2). Additionally, our analysis of APE budgets shows that convective heating dominates APE production (Figure 5) in all the experiments, and therefore may be determinant in the spontaneous TC genesis.

## 5 Sensitivity of spontaneous TC genesis without radiative and surface flux feedbacks

TCs self-emerge without radiative and surface flux feedbacks in 50-day-long simulations in a variety of configurations where we change radiation scheme, microphysics scheme, sea-surface temperature and resolution, while otherwise maintaining the HomoAll configuration. Movie S2 shows the time evolution of surface wind speed of the sensitivity experiments shown in Table 2. Regarding model physics, we observe clearly defined TCs by day 14 when using the Thompson Microphysics scheme, and by day 24 when using RRTM radiation. In the SST sensitivity experiments, TCs first appear by day 14 with a surface temperature of 305K and by day 21 with surface temperature of 297K. When we set the Coriolis parameter to $f = 3 \times 10^{-4}$ s$^{-1}$, TCs first appear by day 33, and when $f = 1 \times 10^{-4}$ s$^{-1}$, convection remains randomly distributed and TCs do not emerge even when running the simulation to 100 days (not shown). This is consistent with the results of Muller & Romps (2018), who showed that spontaneous TC genesis does not occur when $f = 1 \times 10^{-4}$ s$^{-1}$ in a similar computation domain. In Movie S3, spontaneous TC genesis occurs by day 16 when the horizontal grid-spacing is reduced to 1km in the HomoAll configuration.

Figure 6 shows the minimum surface pressure and maximum surface wind speed in the domain for the sensitivity experiments shown in Table 2. We observe maximum surface wind speeds greater than 20ms$^{-1}$ and decreasing minimum surface pressure in all of the simulation except with $f = 1 \times 10^{-4}$ s$^{-1}$, where the wind speed and surface pressure remain relatively flat, consistent with the random convection shown in Movie S2. It is important to note that the minimum surface pressure shows a decreasing trend accompanied by the increase in maximum surface wind speed by the end of the simulation period in the sensitivity experiments for radiation physics, $f = 3 \times 10^{-4}$ s$^{-1}$, and the experiment with surface temperature of 297K, suggesting that the TCs in these sensitivity experiments have not yet finished intensifying. However, this trend is not observed in the experiment with $f = 1 \times 10^{-4}$ s$^{-1}$. In the experiment with 1km grid-spacing, TCs reach similar intensities as those of the experiments in Table 1 by day 30 (Movie S3). The sensitivity study shows that our results are robust over different choices of model physics.

Here, we speculate on potential explanations for the dependence of spontaneous TC genesis to the value of $f$. The spatial scale of TCs may be approximately proportional to $1/f$ (Zhou et al. 2013; Chavas and Emanuel 2014), so TCs in the simulations with $f = 1 \times 10^{-4} \text{ s}^{-1}$ would have 3 times larger spatial scale than in the case with $f = 3 \times 10^{-4} \text{ s}^{-1}$ if all other factors are equal. Therefore, using the same computing domain, the $f = 1 \times 10^{-4} \text{ s}^{-1}$ simulation may not be able to accommodate a TC. Another plausible explanation is that increasing the rotation rate reduces the scale separation between convective organization and the deformation radius (R = NH/$f$ where N is the Brunt-Väisälä frequency and H the height of the tropopause), favoring TC genesis (Ooyama 1982). The deformation radius is roughly the minimum spatial scale that is affected by the rotation and is often much larger than the scale of individual convective storms. Increasing rotation rate would reduce the deformation radius and thereby the scale separation between convective storms and the deformation radius. This makes it easier for organized convection to be affected by the planetary rotation, favoring TC genesis. In Table 3 we show the value of the deformation radius. We compute the deformation radius using the mean sounding of the last 50 days of the simulations shown in Table 1. We observe that in contrast to the cases with $f = 5 \times 10^{-4} \text{ s}^{-1}$ and $f = 3 \times 10^{-4} \text{ s}^{-1}$, the deformation radius of the case with $f = 1 \times 10^{-4} \text{ s}^{-1}$ becomes larger than the simulated domain in all the cases of Table 1, making this a plausible explanation for why we do not observe spontaneous TC genesis in this case. A detailed investigation of the hypotheses requires large-domain simulations and is left for future work.

## 6 Main findings and implications

This paper, for the first time, shows that spontaneous TC genesis can occur after turning off both radiative and surface-flux feedbacks in $f$-plane CRM simulations. The simulated TCs in all of our experiments have realistic horizontal and vertical structures, and our simulation results are robust to varying horizontal resolutions, initial conditions, and a wide range of model physics. This result challenges our previous understanding that spontaneous TC genesis requires either radiative or surface-flux feedbacks (Wing et al. 2018; Muller and Romps 2018).

We find that a high Coriolis parameter favors spontaneous TC genesis. In our computing domain of 1024 km × 1024 km, the minimum $f$ that allows TCs to self-emerge in HomoAll is $f = 3 \times 10^{-4}\,\mathrm{s^{-1}}$, which has been used before in studying spontaneous TC genesis. A plausible explanation for this dependence suggests that our simulation domain may be too small to accommodate spontaneous TC genesis in the HomoAll simulations when the Coriolis parameter becomes smaller. Another explanation suggests that as the Rossby radius of deformation and the convective scales become closer, the probability of convecting regions becoming controlled by the background rotation increases. Other hypothesis considers that a high Coriolis parameter accelerates spontaneous TC genesis by increasing the effect of planetary rotation on the production of vorticity by stretching. Therefore, this dependence of TC genesis on the Coriolis parameter deserves further exploration that consider a larger computation domain or even a different geometry, as recent studies have found that the minimum Coriolis parameter needed for TC genesis depends on the Rhines scale which on an f-plane becomes infinite (Chavas and Reed 2019). In all these further explorations, the APE framework offers an opportunity to re-examine the role of convective heating in the TC genesis process. Therefore, our results contribute to our understanding of the nature of convection and TC genesis and show that, even in earth-like conditions, a cooperative intensification between convective heating and the overturning circulation may contribute to the organization of a TC through its role in APE production, a possibility that had been disregarded in recent studies. Further exploration of the role of convective heating in the production of APE in under Earth-like conditions is desirable to pursue in the future.

Our results are consistent with the *broadly defined* conditional instability of the second kind (CISK), if we define CISK as a cooperative instability between atmospheric flows and convection that does not require surface-flux feedbacks or radiative feedbacks (Bretherton 2003). Conventional CISK studies mainly focused on linear stability analysis or computer simulations with parameterized convection (Ooyama 1982; Smith 1997; Montgomery and Smith 2014). These studies are, therefore, subject to criticisms on assumptions in the representation of dynamics or convection. Their simulated TCs are often with a much smaller spatial scale than that of the observed TCs. To the best of our knowledge, this paper presents the first 3D nonlinear CRM simulations that actually show that TC genesis can result from interactions between convection and atmospheric circulations. This result is, therefore, a significant advancement in our understanding of TC genesis.

Can convection drive large-scale circulations in the absence of radiative and surface-flux feedbacks (Ooyama 1982; e.g. Emanuel et al. 1994)? This is a central question in tropical atmospheric dynamics. This paper and recent research show that cooperative interactions between convection and atmospheric circulations can lead to a wide spectrum of convectively coupled circulations, including convective self-aggregation (Muller and Bony 2015; Yang 2018a, 2019, 2021), TCs (Ooyama 1982; Montgomery and Smith 2014), convectively coupled equatorial waves (Mapes 2000; Kuang 2008; Andersen and Kuang 2008) and the Madden-Julian Oscillation (MJO) (Yang and Ingersoll 2013, 2014; Wang et al. 2016). These studies suggest that convection can indeed drive large-scale circulations without radiative and surface-flux feedbacks.

We use an APE-centric framework (Yang 2018a), which complements the widely used MSE analysis. Our APE analyses show that convective heating coincides with positive buoyancy anomalies (Figure 4) and dominates APE production during both the genesis and mature stages of TC development (Figure 5). The fact that convection dominates the APE production even in the full-physics simulation may help explain why spontaneous TC genesis can exist without the radiative and surface-flux feedbacks. The computation of APE requires defining the atmosphere state of minimal potential energy after adiabatic rearrangement of air mass. The minimal potential energy state may be sensitive to the choice of methods. This challenge is particularly notable when accounting for the phase changes of atmospheric water vapor—a moist APE framework (Lorenz 1978; Randall and Wang 1992; Wong et al. 2016). It is then important to note that we use a "dry" APE framework, which treats convective heating as an external heat source. In future studies, it is desirable to analyze our experiments using "moist" variables, including the MSE and moist APE, which may provide additional insights on the spontaneous TC genesis. Aside from an approach centered in thermodynamics, future work focusing on the dynamics (*e.g.*, analysis of vorticity) of spontaneous TC genesis is necessary. The dynamic and thermodynamic approaches are complementary, and a complete picture of the TC genesis should consider both.

The energy that powers TCs ultimately comes from the ocean (Emanuel 1986, 2003), which transfers energy to the atmosphere primarily through surface sensible and latent heat fluxes. Our experiments are not exceptions, as the surface energy fluxes are key to sustain a moist convecting atmosphere. However, our experiments show that when radiative or surface-flux feedbacks are not active, convective heating may be capable of producing horizontal pressure perturbations, allowing spontaneous TC genesis to occur.

This paper aims to understand TC genesis by studying it in RRCE. In RRCE simulations, TCs can self-emerge, but this process often takes $10 - 30$ days (Wing et al. 2016; Emanuel 2018; Muller and Romps 2018; Carstens and Wing 2020). This timescale is long comparing to that of synoptic-scale disturbances in the tropical atmosphere: physical processes leading to TC genesis in RRCE might be less efficient than synoptic-scale weather systems. Therefore, there are likely other physical processes that promote TC genesis in the real atmosphere. In future studies, it would be useful to repeat our simulations using a hierarchy of numerical models, which may include aqua-planet GCMs with uniform SSTs, aqua-planet GCMs with realistic SST distributions, and GCMs with realistic topography and SST distributions. This approach will not only test the robustness of our results but will also help bridge the gap between highly idealized studies and observation-based studies.

**Appendix 1**

Sensitivity of Spontaneous TC genesis and APE evolution to relaxation of mean wind speed.

Here we show the evolution of four 50-days simulations with no horizontal wind speed nudging. The simulations are otherwise identical to Control, HomoRad, HomoSfc and HomoAll. Spontaneous TC genesis occurs in the Control, HomoRad, HomoSfc and HomoAll configurations without relaxing the mean wind to zero at all levels. In Figure A1 and Movie S4 we show map views of the surface wind speed and surface pressure in simulations where we do not relax the mean wind speed to zero. The simulations are qualitatively similar to their counterparts with wind speed nudging, showing various vortices with maximum wind speeds around 50 ms$^{-1}$. Figure A2 shows that the evolution of maximum wind speed and minimum surface pressure follows closely those of Control, HomoRad, HomoSfc and HomoAll. Figure A3 shows that the APE and APE budgets are also qualitatively similar to the experiments with wind speed nudging for the first 50 days of the simulations.

## Data availability

The namelist files to run the experiments and the analysis code is available at https://github.com/aramirezreyes/RamirezReyes_Yang_2020_SpontaneousCyclogenesis

## Acknowledgments

This work was supported by Laboratory Directed Research and Development (LDRD) funding from Berkeley Lab, provided by the Director, Office of Science, of the U.S. Department of Energy under contract DE-AC02-05CH11231. Argel Ramírez Reyes' doctoral studies are also supported by Mexico's National Council for Science and Technology (CONACYT) and The University of California Institute for Mexico and the United States (UC Mexus) through the CONACYT-UC Mexus doctoral fellowship. Computational resources used were provided by the Department of Energy's National Energy Research Scientific Computing Center (NERSC) at Lawrence Berkeley National Laboratory. The computational model was kindly provided by M. Khairoutdinov and can be obtained through http://rossby.msrc.sunysb.edu/~marat/SAM.html. We thank T. Cronin, S. D. Seidel, D. Chavas and W. Zhou for helpful comments during the early stage of this work, D. Nolan for a helpful discussion at the AGU 2019 Fall Meeting.

**Figures**

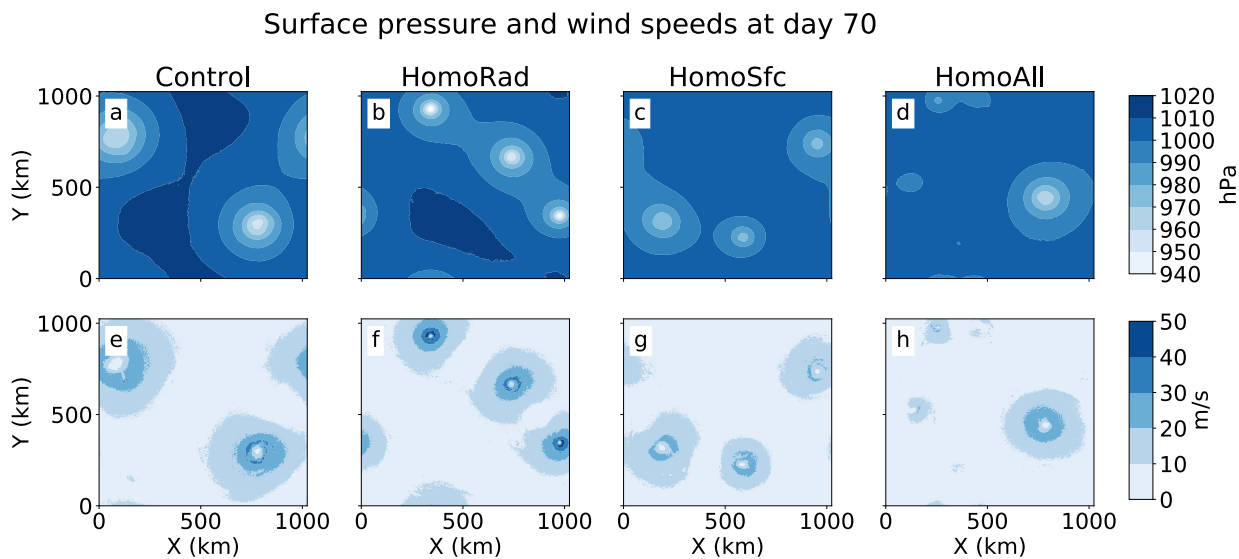

**Figure 1.** Map views of surface pressure and surface wind speed at *t* = 70 days (snapshot). (a-d) contours of surface pressure (hPa), (e-h) and surface wind speed (m/s). The first, second, third and fourth columns correspond to the Control, HomoRad, HomoSfc and HomoAll simulations with 2-km grid spacing, respectively.

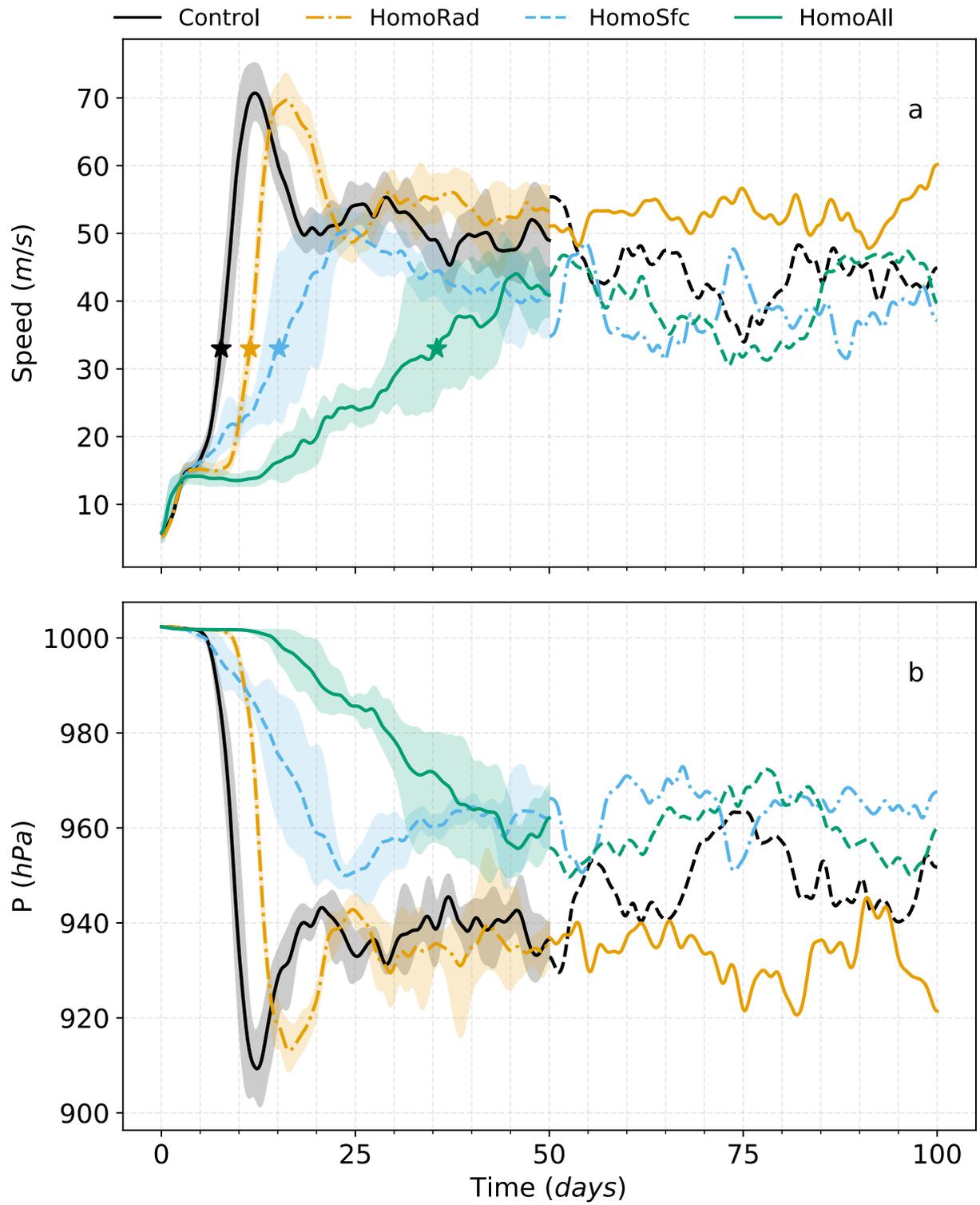

**Figure 2.** Time evolution of maximum surface wind speed (m/s) (a) and minimum surface (hPa) (b). Hourly data is smoothed with a moving average filter with window = 20 hours.

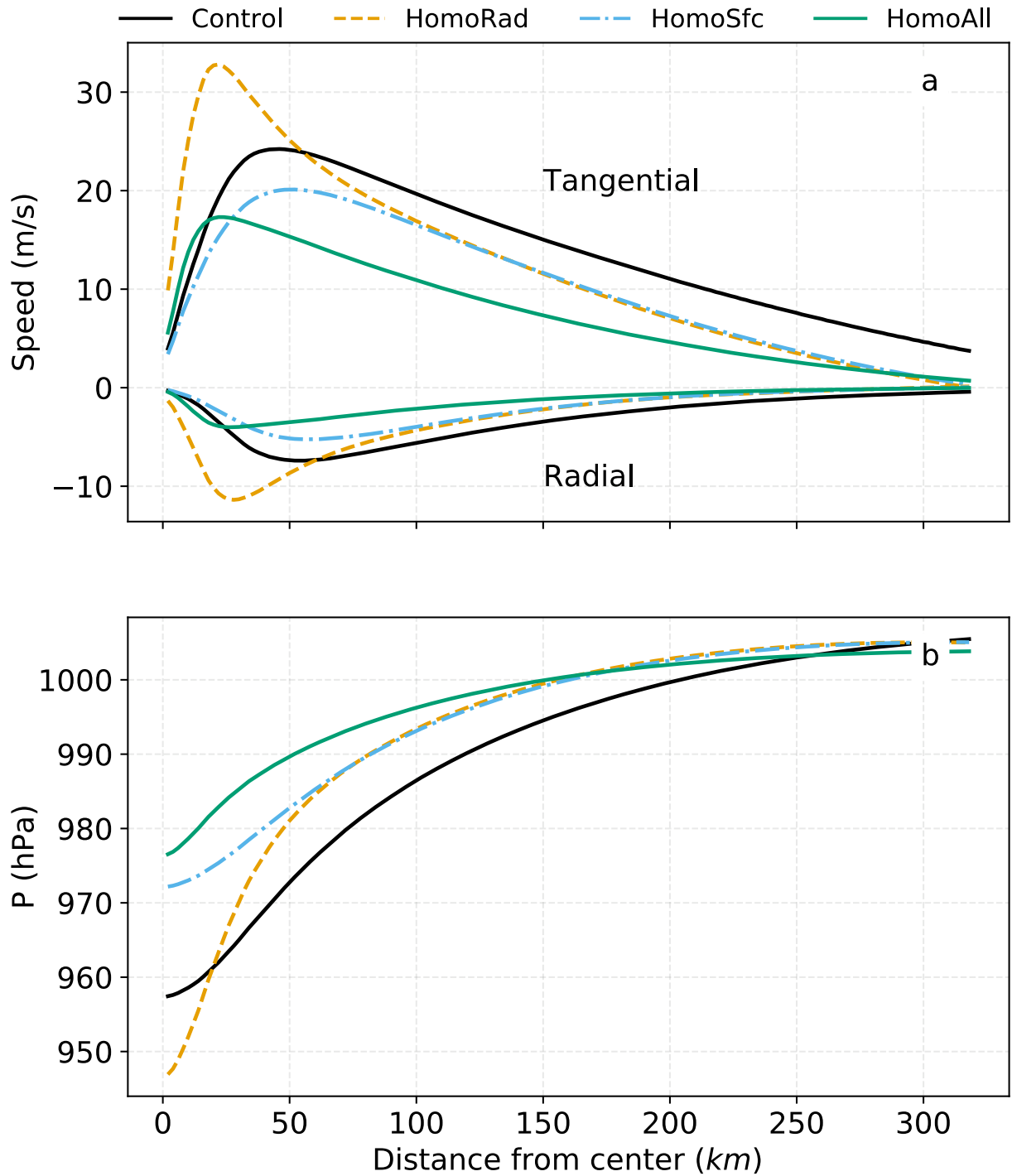

**Figure 3.** Azimuthal averages of radial and tangential wind speed at the surface and surface pressure in the composite TCs. (a) Radial and tangential wind speed (m/s). (b) Surface pressure (hPa).

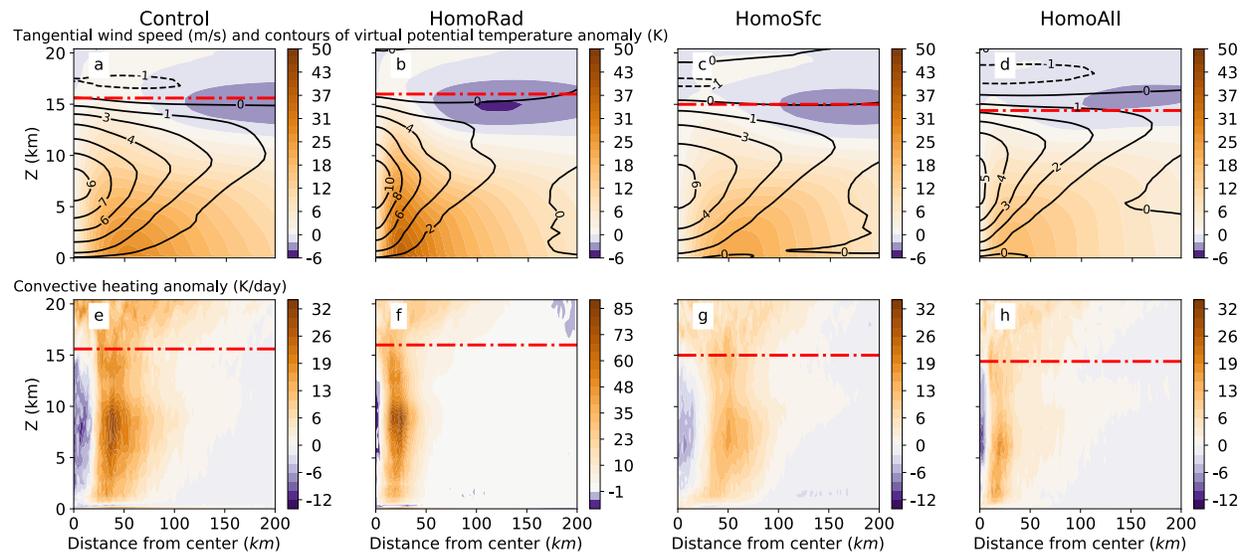

**Figure 4.** Azimuthal average of tangential wind speed, temperature anomaly and convective heating rate in the composite TC. (a-d) colored contours of tangential wind speed (m/s) and black contours of virtual potential temperature anomaly (K). (e-h) colored contours of convective heating rate anomaly (K/day) and a red line indicating the height of the radiative tropopause defined by zero radiative cooling rate. The first, second, third and fourth columns correspond to the Control, HomoRad, HomoSfc and HomoAll simulations, respectively.

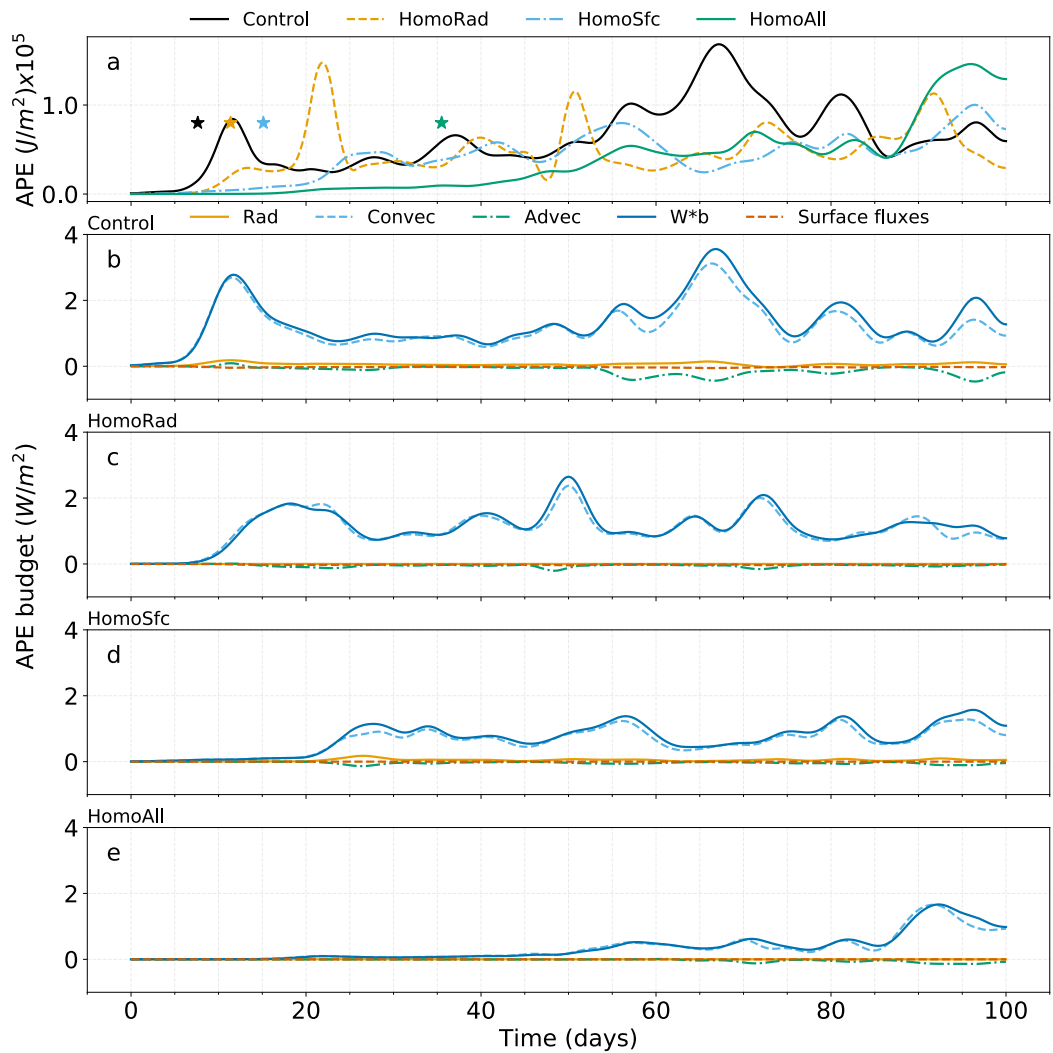

**Figure 5.** The available potential energy (APE) and its budget, integrated from the surface to the radiative tropopause . (a) The APE in the Control, HomoRad, HomoSfc and HomoAll experiments. (b) The APE budgets for the Control simulation. (c) The APE budgets for the HomoRad simulation. (d) The APE budgets for the HomoSfc simulation. (e) The APE budgets for the HomoAll simulation. In (b-e) orange, solid line represents the radiation term, light blue, dashed, represents the convection term, green with dashes and dots represents the advection term, dark blue and solid line represents the conversion to kinetic energy, and red, dashed

represents the surface fluxes term. Data is smoothed with a moving average filter with window width = 20 hours.

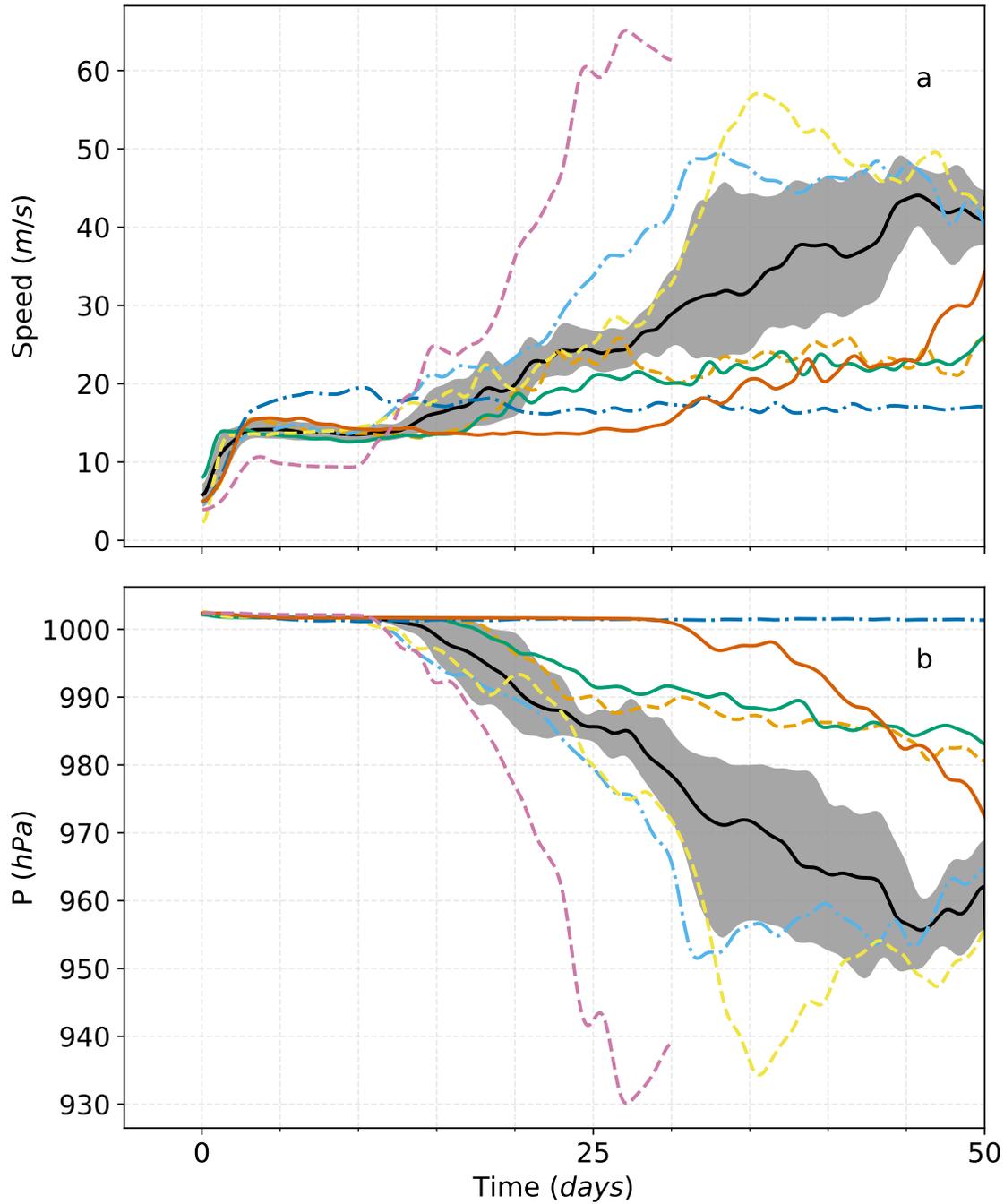

**Figure 6.** Time evolution of maximum surface wind speed and minimum surface pressure for simulations at 2-km grid-spacing. (a) Shows wind speed (m s⁻¹), (b) shows minimum surface pressure (hPa). Different lines correspond to different experiments (see legend). Hourly data is smoothed with a moving average filter with window = 20 points.

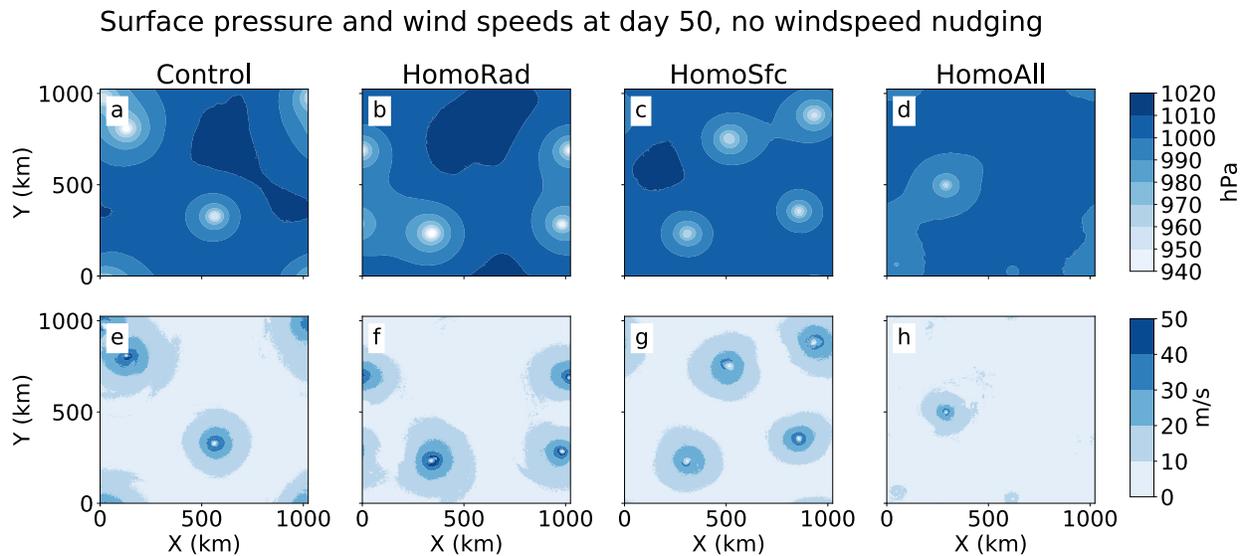

**Figure A1.** Map views of surface pressure and surface wind speed at $t = 50$ days (snapshot). (a-d) contours of surface pressure (hPa), (e-h) and surface wind speed (m/s). The first, second, third and fourth columns correspond to the Control, HomoRad, HomoSfc and HomoAll simulations with 2-km grid spacing without mean wind speed nudging.

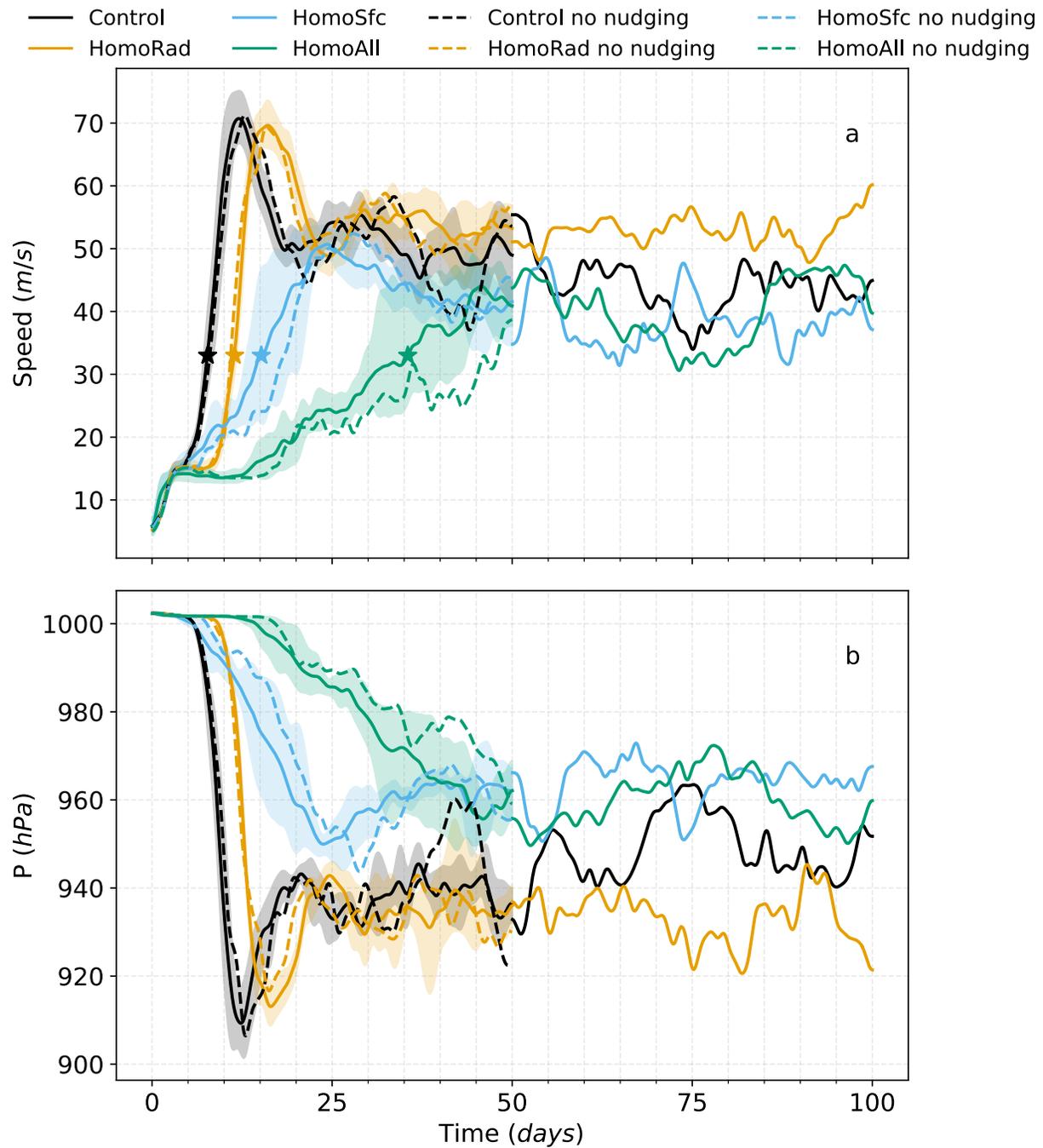

**Figure A2.** Time evolution of maximum surface wind speed and minimum surface pressure for simulations at 2-km grid-spacing. (a) Shows wind speed (m s$^{-1}$), (b) shows minimum surface pressure (hPa). Different lines correspond to different experiments (see legend). Continuous lines show values in the simulations with mean wind speed nudging and dashed lines show values in the simulations without mean wind speed nudging. Hourly data is smoothed with a moving average filter with window = 20 hours.

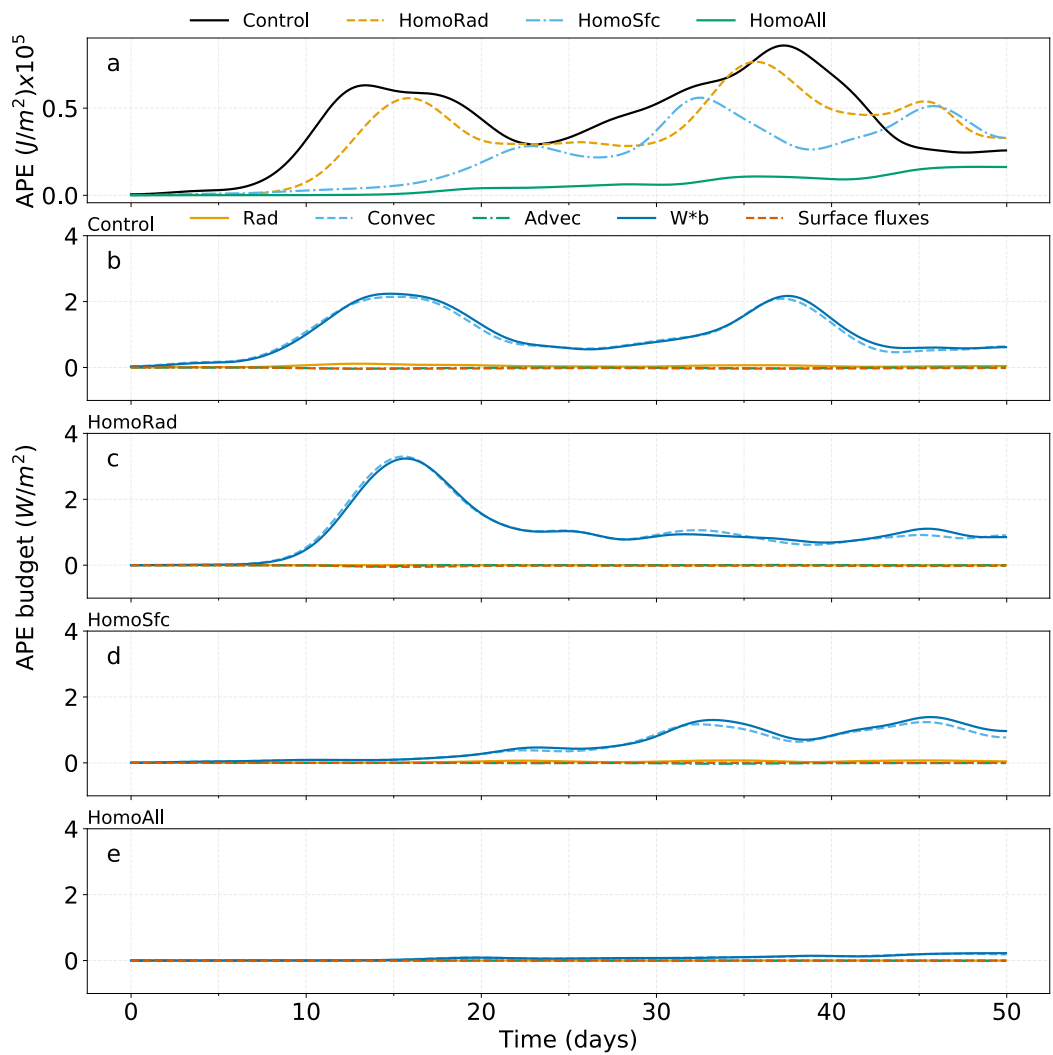

**Figure A3.** The available potential energy (APE) and its budget, integrated from the surface to the radiative tropopause for 50-day experiments without relaxation of horizontal mean wind. (a) The APE in the Control, HomoRad, HomoSfc and HomoAll experiments. (b) The APE budgets for the Control simulation. (c) The APE budgets for the HomoRad simulation. (d) The APE budgets for the HomoSfc simulation. (e) The APE budgets for the HomoAll simulation. In (b-e)

orange, solid line represents the radiation term, light blue, dashed, represents the convection term, green with dashes and dots represents the advection term, dark blue and solid line represents the conversion to kinetic energy, and red, dashed represents the surface fluxes term. Data is smoothed with a moving average filter with window width = 20 hours.

| Name | Feedbacks Removed | Grid size | Domain size | Simulation length |
|------|-------------------|-----------|-------------|-------------------|
| Control | None | 512x512x80 | 1000 km x 1000 km x 34.8 km | 100 days |
| HomoSfc | Surface-flux feedbacks removed | 512x512x80 | 1000 km x 1000 km x 34.8 km | 100 days |
| HomoRad | Radiative feedbacks removed | 512x512x80 | 1000 km x 1000 km x 34.8 km | 100 days |
| HomoAll | Radiative and surface-flux feedbacks removed | 512x512x80 | 1000 km x 1000 km x 34.8 km | 100 days |

**Table 1.** Summary of simulation parameters in the mechanism-denial experiments.

| | Parameter changed | Base configuration | Description of change | Simulation length |
|---|-------------------|--------------------|-----------------------|-------------------|
| 1 | Sea surface temperature | HomoAll | 297 K | 50 days |
| 2 | Sea surface temperature | HomoAll | 305 K | 50 days |

| 3 | Initial conditions | Control, HomoSfc, HomoRad and HomoAll | Changed initialization profile | 50 days |
|---|---|---|---|---|
| 4 | Coriolis parameter | HomoAll | $f = 1 \times 10^{-4} \text{ s}^{-1}$ | 50 days |
| 5 | Coriolis parameter | HomoAll | $f = 3 \times 10^{-4} \text{ s}^{-1}$ | 50 days |
| 6 | Radiation scheme | HomoAll | Used RRTM radiation model | 50 days |
| 7 | Microphysics scheme | HomoAll | Used Thompson microphysics scheme | 50 days |
| 8 | Resolution | HomoAll | Grid spacing of 1 km | 30 days |

**Table 2.** Summary of simulation parameters in the sensitivity experiments.

| Deformation radius = NH/$f$ (km) | Control | HomoRad | HomoSfc | HomoAll |
|---|---|---|---|---|
| $f = 4.97 \times 10^{-4} \text{ s}^{-1}$ | **386.1** | **397.1** | **335.7** | **399.0** |
| $f = 3 \times 10^{-4} \text{ s}^{-1}$ | 639.6 | 657.8 | 556.1 | 509.4 697.6* |
| $f = 1 \times 10^{-4} \text{ s}^{-1}$ | 1919.0 | 1973.5 | 1668.4 | 1528.2 1957.9* |
|  |  |  |  |  |

**Table 3.** Deformation radius for different values of the Coriolis parameter using the mean sounding of last 50 days of the simulations in Table 1 to compute N (the Brunt-Väisälä frequency) and H (the height of the tropopause). The values marked with an asterisk are the deformation radii computed using the mean sounding from the sensitivity simulations with different $f$ (see Table 2).